\documentclass[draft]{agujournal2019}
\usepackage{url} %this package should fix any errors with URLs in refs.
\usepackage{booktabs}   % 三线表核心包
\usepackage{multirow}   % 多行合并
\usepackage{hyperref}
\usepackage[inline]{trackchanges} %for better track changes. finalnew option will compile document with changes incorporated.
\usepackage{soul}
\draftfalse

\journalname{Geophysical Research Letters}
\usepackage{amsmath}
\usepackage{natbib} 
\begin{document}
%%%%%%%%%%%%%%%%%%%%%%%%%%%%%%%%%%%%%%%%%%%%%%%
%  TITLE
%
% (A title should be specific, informative, and brief. Use
% abbreviations only if they are defined in the abstract. Titles that
% start with general keywords then specific terms are optimized in
% searches)
%
%%%%%%%%%%%%%%%%%

% Example: \title{This is a test title}

\title{Meteorologically-Informed Adaptive Conformal Prediction for Tropical Cyclone Intensity Forecasting}

%%%%%%%%%%%%%%%%%%%%%%%%%%%%%%%%%%%%%%%%%%%%%%%
%
%  AUTHORS AND AFFILIATIONS
%
%%%%%%%%%%%%%%%%%%%%%%%%%%%%%%%%%%%%%%%%%%%%%%%

% Authors are individuals who have significantly contributed to the
% research and preparation of the article. Group authors are allowed, if
% each author in the group is separately identified in an appendix.)

% List authors by first name or initial followed by last name and
% separated by commas. Use \affil{} to number affiliations, and
% \thanks{} for author notes.
% Additional author notes should be indicated with \thanks{} (for
% example, for current addresses).

% Example: \authors{A. B. Author\affil{1}\thanks{Current address, Antartica}, B. C. Author\affil{2,3}, and D. E.
% Author\affil{3,4}\thanks{Also funded by Monsanto.}}

\authors{Xuepeng Chen\affil{1}, Jing-Jia Luo\affil{2,3}Qingqing Li\affil{3,4,5},Fan Meng\affil{1}}

% \affiliation{1}{First Affiliation}
% \affiliation{2}{Second Affiliation}
% \affiliation{3}{Third Affiliation}
% \affiliation{4}{Fourth Affiliation}
\affiliation{1}{School of Artificial Intelligence ,Nanjing University of Information Science and Technology, Nanjing, China}
\affiliation{2}{Institute for Climate and Application Research (ICAR), Nanjing University of Information Science and Technology, Nanjing 210044, China}
\affiliation{3}{Key Laboratory of Meteorological Disaster, Ministry of Education (KLME), Joint International Research Laboratory of Climate and Environment Change (ILCEC), Collaborative Innovation Center on Forecast and Evaluation of Meteorological Disasters (CIC-FEMD), Nanjing University of Information Science \& Technology, Nanjing 210044, China}
\affiliation{4}{School of Atmospheric Sciences, Nanjing University of Information Science and Technology, Nanjing, China}
\affiliation{5}{Pacific Typhoon Research Center and Key Laboratory of Meteorological Disaster of the Ministry of Education, Nanjing University of Information Science and Technology, Nanjing, China}

%(repeat as many times as is necessary)

% Corresponding author mailing address and e-mail address:

% (include name and email addresses of the corresponding author.  More
% than one corresponding author is allowed in this LaTeX file and for
% publication; but only one corresponding author is allowed in our
% editorial system.)

% Example: \correspondingauthor{First and Last Name}{email@address.edu}

\correspondingauthor{Fan Meng}{meng@nuist.edu.cn}

%%%%%%%%%%%%%%%%%%%%%%%%%%%%%%%%%%%%%%%%%%%%%%%
% KEY POINTS
%%%%%%%%%%%%%%%%%%%%%%%%%%%%%%%%%%%%%%%%%%%%%%%
%  List up to three key points (at least one is required)
%  Key Points summarize the main points and conclusions of the article
%  Each must be 140 characters or fewer with no special characters or punctuation and must be complete sentences

% Example:
% \begin{keypoints}
% \item	List up to three key points (at least one is required)
% \item	Key Points summarize the main points and conclusions of the article
% \item	Each must be 140 characters or fewer with no special characters or punctuation and must be complete sentences
% \end{keypoints}

\begin{keypoints}
\item A physics-informed conformal prediction method is proposed for tropical cyclone intensity uncertainty forecasting.
\item The method integrates confidence levels with the cyclone intensity stages to adaptively optimize the reliability of the interval.
\item Extensive experiments demonstrate that our model significantly enhances prediction reliability during rapid intensification phases.  

\end{keypoints}

%%%%%%%%%%%%%%%%%%%%%%%%%%%%%%%%%%%%%%%%%%%%%%%
%
%  ABSTRACT and PLAIN LANGUAGE SUMMARY
%
% A good Abstract will begin with a short description of the problem
% being addressed, briefly describe the new data or analyses, then
% briefly states the main conclusion(s) and how they are supported and
% uncertainties.

% The Plain Language Summary should be written for a broad audience,
% including journalists and the science-interested public, that will not have 
% a background in your field.
%
% A Plain Language Summary is required in GRL, JGR: Planets, JGR: Biogeosciences,
% JGR: Oceans, G-Cubed, Reviews of Geophysics, and JAMES.
% see http://sharingscience.agu.org/creating-plain-language-summary/)
%
%%%%%%%%%%%%%%%%%%%%%%%%%%%%%%%%%%%%%%%%%%%%%%%

%% \begin{abstract} starts the second page

\begin{abstract}
Rapid intensification (RI) of tropical cyclones (TCs) poses a great challenge due to their highly nonlinear dynamics and inherent uncertainties. Conventional statistical dynamics and artificial intelligence prediction models typically rely on static parameterization schemes, which limits their ability to capture the non-stationary error structure in the intensity evolution. To address this issue, we propose a physically-inspired covariate adaptive conformal prediction framework that dynamically adjusts uncertainty quantification by incorporating process information such as intensity and evolutionary stage. Our approach not only surpasses state-of-the-art models in point prediction accuracy, but also delivers physically consistent and interpretable forecast intervals, establishing a more process-aware framework for probabilistic prediction of extreme weather events.
\end{abstract}

\section*{Plain Language Summary}
Rapid intensification (RI) of tropical cyclones is an important challenge in weather forecasting. This process involves complex physical interactions and a high degree of uncertainty, especially when the storm intensity is rapidly increasing. However, most of the current uncertainty quantification methods use static setups that cannot effectively adapt to changing storm conditions. As a result, forecasts may lack accuracy or provide overly conservative warnings. In this study, we propose a new method that dynamically adjusts the uncertainty range of forecasts based on the stage of storm development and forecast timeliness. Inspired by the storm evolution process, our method allows the model to respond more accurately to different conditions, e.g., expanding the uncertainty range during the RI phase and decreasing it appropriately when stabilizing, thus improving the reliability and usefulness of the forecast. By applying the method to several real-world cases, we demonstrate that it enhances the ability of the forecasting system to handle RI conditions. This adaptive framework provides a new direction for probabilistic forecasting of extreme weather. Given the increasing need for reliable and actionable forecasts in the face of climate change, our framework holds broad relevance for disaster preparedness, emergency response, and policy making.
%%%%%%%%%%%%%%%%%%%%%%%%%%%%%%%%%%%%%%%%%%%%%%%
%
%  BODY TEXT
%
%%%%%%%%%%%%%%%%%%%%%%%%%%%%%%%%%%%%%%%%%%%%%%%

%%% Suggested section heads:
 \section{Introduction}
TCs are among the most destructive natural disasters worldwide. Their intensity variations, especially the process of RI, pose significant challenges for early warning and disaster preparedness \citep{manikanta2023recent}. While the accuracy of track forecasting has improved substantially over the past few decades, intensity forecasting—particularly for RI events—remains a major challenge in TC research \citep{heming2019review,wang2023review}.

Traditional statistical models such as SHIPS and LGEM are widely used in operational forecasting \citep{demaria2021operational}. However, due to their relatively simple structures, they struggle to capture the nonlinear characteristics inherent in cyclone intensity changes \citep{wei2021advanced}. In recent years, Artificial intelligence techniques have gained increasing traction in TC forecasting, demonstrating strong modeling capabilities, particularly in intensity prediction \citep{chen2020machine}. \cite{xu2021deep} employed deep neural networks to significantly enhance short-term intensity forecast accuracy, while \cite{wang2021tropical} utilized convolutional neural networks to process hurricane imagery, achieving accurate estimation of intensity.

Beyond point forecasts, researchers have begun to focus on quantifying predictive uncertainty \citep{rozoff2011new,demaria2009new}. \cite{meng2023tropical} developed a hybrid machine learning framework that outputs both predictive values and confidence intervals, highlighting the potential of probabilistic forecasting. However, such models typically rely on strong distributional assumptions that are often difficult to sustain in complex natural environments, especially meteorological forecasting tasks \citep{duan2020ngboost,kang2016bayesian,lakara2021evaluating}.

Against this backdrop,Conformal Prediction(CP), a non-parametric, distribution-free method for constructing predictive intervals, has gained significant attention due to its theoretical guarantees on coverage \citep{zhou2024conformal}. \cite{angelopoulos2021gentle} highlighted CP’s ability to generate valid prediction intervals without distributional assumptions in both regression and classification tasks.  CP has been successfully applied in financial market forecasting \citep{bastos2024conformal} and energy marketing predictions \citep{kath2021conformal}.  \cite{meng2024uncertainty} also applied CP to TC track forecasting, demonstrating its promising accuracy and adaptability.

However, existing CP methodologies typically adopt a fixed confidence level, failing to account for the dynamic and heterogeneous uncertainty structures that characterize different stages of cyclone evolution. As noted by \cite{kaplan2010revised}, the RI phase is marked by elevated forecast uncertainty, suggesting that a static treatment of confidence levels may result in intervals that are overly conservative or insufficiently informative. 

To address these challenges, we introduce a physics-informed adaptive CP framework, which integrates cyclone-specific physical indicators, such as instantaneous intensity and intensity tendency, into the dynamic calibration of confidence levels. This is, to our knowledge, the first work to systematically integrate CP, physical stage awareness in tropical cyclone forecasting. Beyond technical contributions, our method has significant implications for early warning systems, particularly in high-risk stages like rapid intensification, providing a new paradigm for risk-aware and adaptive forecast models in operational meteorology.
%
% The main text should start with an introduction. Except for short
% manuscripts (such as comments and replies), the text should be divided
% into sections, each with its own heading.

% Headings should be sentence fragments and do not begin with a
% lowercase letter or number. Examples of good headings are:

 \section{ Data and Methods}
\subsection{ Data  }
The data used in this study is from SHIPS (Statistical Hurricane Intensity Prediction Scheme) dataset , a statistical-dynamical model operated by the NHC. The SHIPS model integrates a variety of meteorological and oceanographic variables derived from operational forecast models, satellite observations, and reanalysis data \citep{demaria1999updated}.  

We combined data from three major ocean basins: the Atlantic(AL), Central Pacific, and Eastern North Pacific(EP), ensuring a broad geographical representation of TC development under different climatic conditions. It records TC's characteristics at 6-hour intervals, capturing the evolution of storms over time. The combined dataset comprises many variables characterizing TC intensity evolution and environmental conditions. Several key variables in the dataset have been widely recognized as important predictors of TC intensity, including sea surface temperature, vertical wind shear, and latitude—factors critical to TC genesis and intensification \citep{emanuel1987dependence,demaria1994sea} . 

Before training our models, missing values are handled by different ways. Non-feature columns and variables with more than 50\% missing values were excluded. Remaining missing values were addressed via K-nearest neighbors (KNN) imputation for moderately missing features (10–50\% missingness) \citep{jordanov2018classifiers} and Mean filling for features with less than 10\% missingness \citep{emmanuel2021survey}.We used 121 predictor variables after selection and preprocessing from the SHIPS database. The data division in this paper utilizes a stratification strategy based on storm independence and temporal order. Detailed information of these predictor variables and data division can be seen in Table S1 and Text S1 in Supporting Information(SI). 

\begin{figure}
    \centering
    \includegraphics[width=1\linewidth]{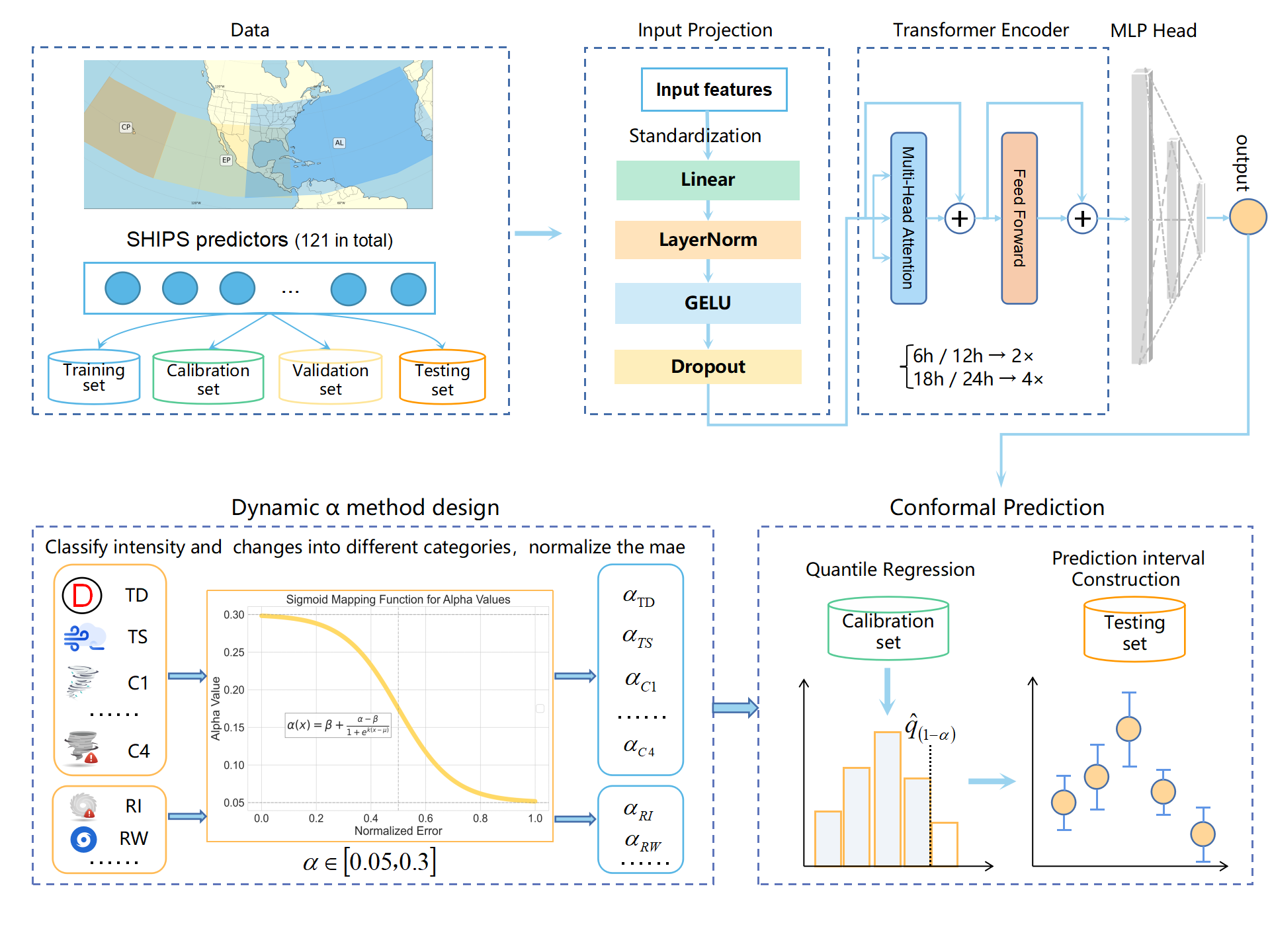}
   \caption{Schematic of the Adaptive CP Framework. During quantile regression, the CP method employs a dynamic $\alpha$ methods that first identifies the current state by determining the intensity category and its change rate. Based on this assessment, a specific confidence level $\alpha$ is assigned. Subsequently, the method computes the quantile $\hat{q}_{1 - \alpha}$
 of the residuals derived from the calibration set. This quantile is then combined with the deterministic point prediction generated by the Transformer model to construct a comprehensive prediction interval on the testing set. }
    \label{fig:1}
\end{figure}. 
\subsection{Adaptive CP}
This study proposed a hybrid prediction framework based on Transformer and CP. The Transformer model has a self-attention mechanism that gives it a strong ability to represent sequences, enabling it to efficiently capture the long-term dependencies and complex nonlinear evolution characteristics of cyclone development processes\citep{vaswani2017attention,zeng2023transformers}.CP, combined with the transformer, is a statistically rigorous framework that constructs prediction intervals with guaranteed coverage probability  \(1-\alpha\) for unseen data, assuming only exchangeability of samples \citep{shafer2008tutorial}. Unlike traditional methods, CP is model-agnostic, enabling seamless integration with any regression or classification model \citep{lei2018distribution}. 

The modeling framework proposed in this study is shown in  figure\ref{fig:1}. The core innovation lies in the design of the dynamic alpha, which is based on the observation that the error distributions of typhoons of different intensity classes under different prediction timescales show significant differences, and such differences are significantly associated with the physical characteristics of the storm such as the core structure and the peripheral circulation. In addition, the initial intensity of the storm and the 12-hour intensity change (DELV-12) have been previously demonstrated in studies, which are closely related to the subsequent 24-hour intensity change \citep{carrasco2014influence,jiang2012relationship}, further supporting the rationality of the stratification based on the intensity class and intensity change. \cite{saffir1973hurricane} grading was used for the intensity class.The definition of RI proposed by \cite{kaplan2003large}—an increase in maximum sustained winds of at least 30 kt within 24 hours—was adopted in this study. Specifically, we converted this definition to a 12-hour intensification threshold of $\geq$15 kt. Based on this, the framework innovatively establishes a two-module hierarchical-mapping architecture: firstly, the error feature matrix is constructed through the two-dimensional hierarchy of storm category and forecast time, and then the maximum-minimum normalization combined with the Sigmoid function is used to construct a nonlinear parameter mapper to achieve the adaptive optimization of the confidence parameter. By preserving the physical correlation between storm features and forecast errors, the results of quantifying uncertainty are ensured to be interpretable in a meteorological sense, and more details are described in the text S2 of SI.

\subsection{Evaluation metrics }
To comprehensively assess the model’s performance, we evaluate its deterministic accuracy using two standard error metrics—Mean Absolute Error (MAE) and Root Mean Square Error (RMSE)—and its probabilistic forecasting skill via the Continuous Ranked Probability Score (CRPS). Furthermore, to appraise the quality of the prediction intervals, we report the Prediction Interval Coverage Probability (PICP), Prediction Interval Width (PIW), and the Winkler score \citep{winkler1972decision}. Detailed formulations for all metrics are provided in the text S3 in SI.
\section{Results}
\subsection{ Predictive Performance}
\begin{figure}
        \centering
        \includegraphics[width=0.75\linewidth]{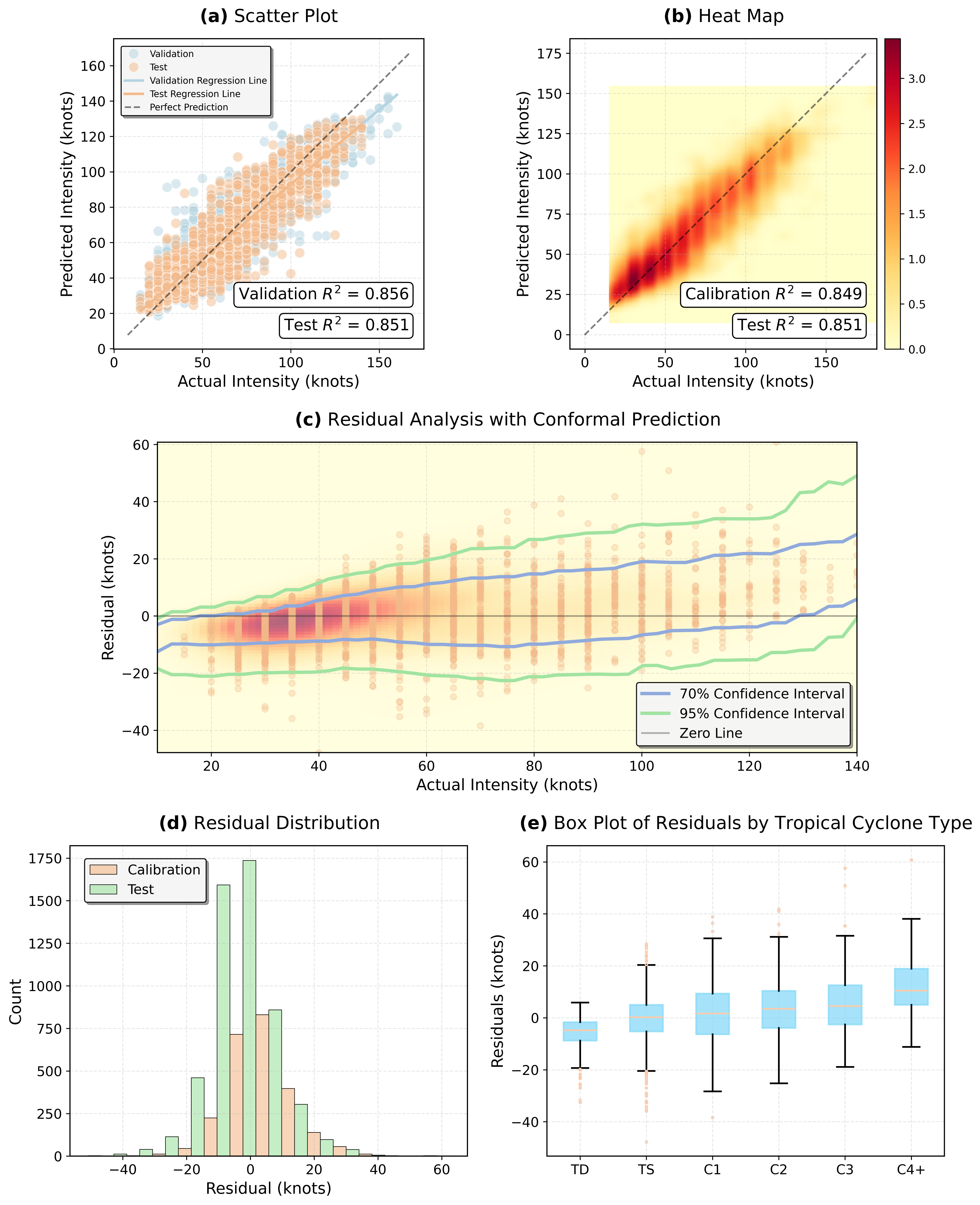}
        \caption{Model performance for 24-hour intensity prediction and residual analysis. (a) Scatter plot comparing predicted and actual TC intensity, with regression lines for validation and test sets. (b) Heat map showing the density of predictions relative to actual values. (c) Residual analysis incorporating CP intervals at 70\% and 95\% confidence levels. (d) Distribution of residuals for calibration and test datasets. (e) Box plot of residuals categorized by TC intensity levels.}
        \label{fig:2}
    \end{figure}

Figure\ref{fig:2} comprehensively shows the performance and uncertainty analysis of the intensity prediction model as follows:(a) Scatterplot shows that the predicted values are highly consistent with the actual values, and the R²s of the validation set and test set are 0.856 and 0.851, respectively, which indicates that the model performs stably and predicts accurately on different datasets.(b) Heat map The predicted density is reflected by the color, and the high-density area is concentrated near the diagonal line, and the R²s of the calibration set and the test set are 0.849 and 0.851, respectively, which further verifies that the prediction effect is good.(c) The analysis of residuals and confidence intervals shows that the prediction errors are larger in high-intensity typhoons, and the 70\% and 95\% confidence intervals become wider with the increase of wind speed, while the errors are smaller and stable in the range of 60-80 knots.(d) The residual distribution approximates a normal shape centered around zero, indicating that there is no systematic bias, and the distributions of the calibration set and the test set are similar, so the model is in good agreement.(e) It's obvious to find the range of residuals is widened as the storm level increases, indicating that strong storms are more challenging to predict.
\begin{figure}
    \centering
    \includegraphics[width=1\linewidth]{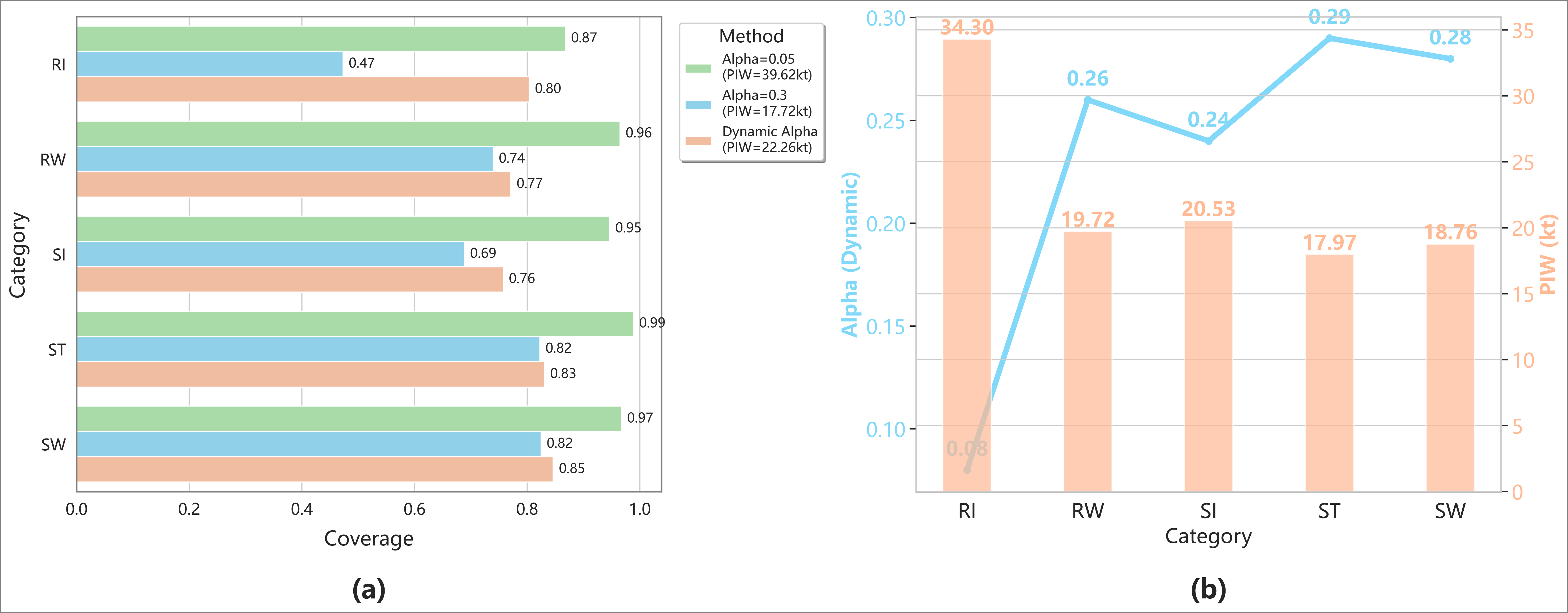}
 \caption{(a) Coverage across intensity change categories under fixed and dynamic CP schemes. (b) Dynamically adjusted alpha and corresponding PIW reflect the method’s adaptability to category-specific uncertainty.  }
    \label{fig:3}
\end{figure}

Figure\ref{fig:3}  illustrates the 18-hour prediction interval coverage and uncertainty performance for different methods . The fixed alpha = 0.05 CP method has the highest coverage in all categories, but the intervals are too wide; the alpha = 0.3 CP method has low coverage. In contrast, the dynamic alpha method achieves a good balance of coverage and interval width in all categories. Especially in the RI category, the coverage of the dynamic method is significantly better than that of alpha = 0.3 (0.80 vs. 0.47), while the interval width is much smaller than that of alpha = 0.05 (34.30 kt vs. 39.62 kt), which demonstrates better adaptability and reliability.

\subsection{Comparison models}
To evaluate the performance of our method, we compare the 24-hour intensity prediction error statistics with three types of models, deterministic point models,probabilistic prediction models and operational models respectively. 
The deterministic point prediction models considered in this study include XGBoost \citep{chen2016xgboost}, LightGBM\citep{ke2017lightgbm}, CatBoost\citep{prokhorenkova2018catboost}, Random Forest\citep{breiman2001random}, and Multi-Layer Perceptron (MLP). To quantify prediction uncertainty, we evaluate several probabilistic models, including Conformalized Quantile Regression (CQR) \citep{romano2019conformalized}, Monte Carlo Dropout \citep{milanes2021monte}, Deep Ensemble \citep{lakshminarayanan2017simple}, DropConnect \citep{mobiny2021dropconnect}, and NGBoost \citep{duan2020ngboost},Specific parameter configurations for these models can be found in text S4 in the SI. For comparison with operational standards, we also include models and consensus forecasts used by the National Hurricane Center (NHC), such as the Decay-Statistical Hurricane Intensity Prediction Scheme (DSHP), the Logistic Growth Equation Model (LGEM) \citep{demaria2009simplified}, the Hurricane Weather Research and Forecast system (HWFI/HWRF), and the official NHC forecasts (OFCL) \citep{simon2018description}, We collected the prediction errors of these models over the EP and AL from 2016 to 2023 (the time span of our test set), and computed the average errors,More detailed results are provided in Table S3 in SI. 
% Requires: \usepackage{graphicx}

\begin{table}[h]
\caption{Comparison of various models on TC intensity forecasting metrics from 2016 to 2023}
\label{tab:1}
    \centering
    \begin{tabular}{llccccccc}
\hline
\textbf{Category} & \textbf{Model} & \textbf{MAE (kt)} & \textbf{RMSE (kt)} & \textbf{R$^2$} & \textbf{CRPS} & \textbf{PICP}& \textbf{PIW} & \textbf{RATIO} \\
\hline
\multirow{5}{*}{Deterministic} 
& XGBoost       & 9.31 & 12.84 & 0.77 & --   & --   & --   & -- \\
& LightGBM      & 9.27 & 12.73 & 0.77 & --   & --   & --   & -- \\
& CatBoost      & 9.33 & 12.82 & 0.77 & --   & --   & --   & -- \\
& RandomForest  & 10.02 & 13.79 & 0.73 & --   & --   & --   & -- \\
& MLP           & 9.42 & 12.96 & 0.76 & --   & --   & --   & -- \\
\multirow{5}{*}{Uncertainty}
& CQR            & 9.71 & 14.10 & 0.77 & --   & 0.91& 44.63 & 49.04 \\
& MC Dropout     & 9.69 & 12.91 & 0.71 & 7.98 & 0.51 & 16.22 & 31.80 \\
& Deep Ensemble  & 9.86 & 12.87 & 0.75 & 8.46 & 0.48 & 15.14 & 31.54 \\
& DropConnect    & 10.34 & 14.29 & 0.72 & 8.86 & 0.40 & 12.31 & 30.78 \\
& NGBoost        & 9.55 & 13.13 & 0.76 & 7.02 & 0.84 & 33.89 & 40.35 \\
\multirow{5}{*}{Operational}
& DSHP           & 10.18  & -- & -- & -- & -- & -- & -- \\
& LGEM           & 10.14  & -- & -- & -- & -- & -- & -- \\
& HWFI           & 9.61  & -- & -- & -- & -- & -- & -- \\
& OFCL           & 8.56 & -- & -- & -- & -- & -- & -- \\
& Our Method &7.84& 10.74& 0.84& -- & 0.81 & 25.33 & 31.27 \\
\hline
\end{tabular}

\end{table}

Table\ref{tab:1}  shows the results of comparing our model with various models, in the comparison of deterministic point prediction, our model achieves the best result from 2016 to 2023 among operational models, meanwhile, our method is better than the deterministic prediction of machine learning models such as lightgbm. In terms of uncertainty prediction, we hope that the higher the coverage the better, and the narrower the interval width the better, in order to better compare, we made the ratio of the two to get the ratio, although the best result is drop connect, but its coverage is lower than the preset 95\% confidence level, and is not taken into account, in this way, our results show a more robust performance, which takes into account both the interval width and guarantees certain coverage.

\subsection{Case Studies}
\begin{figure}
    \centering
    \includegraphics[width=1\linewidth]{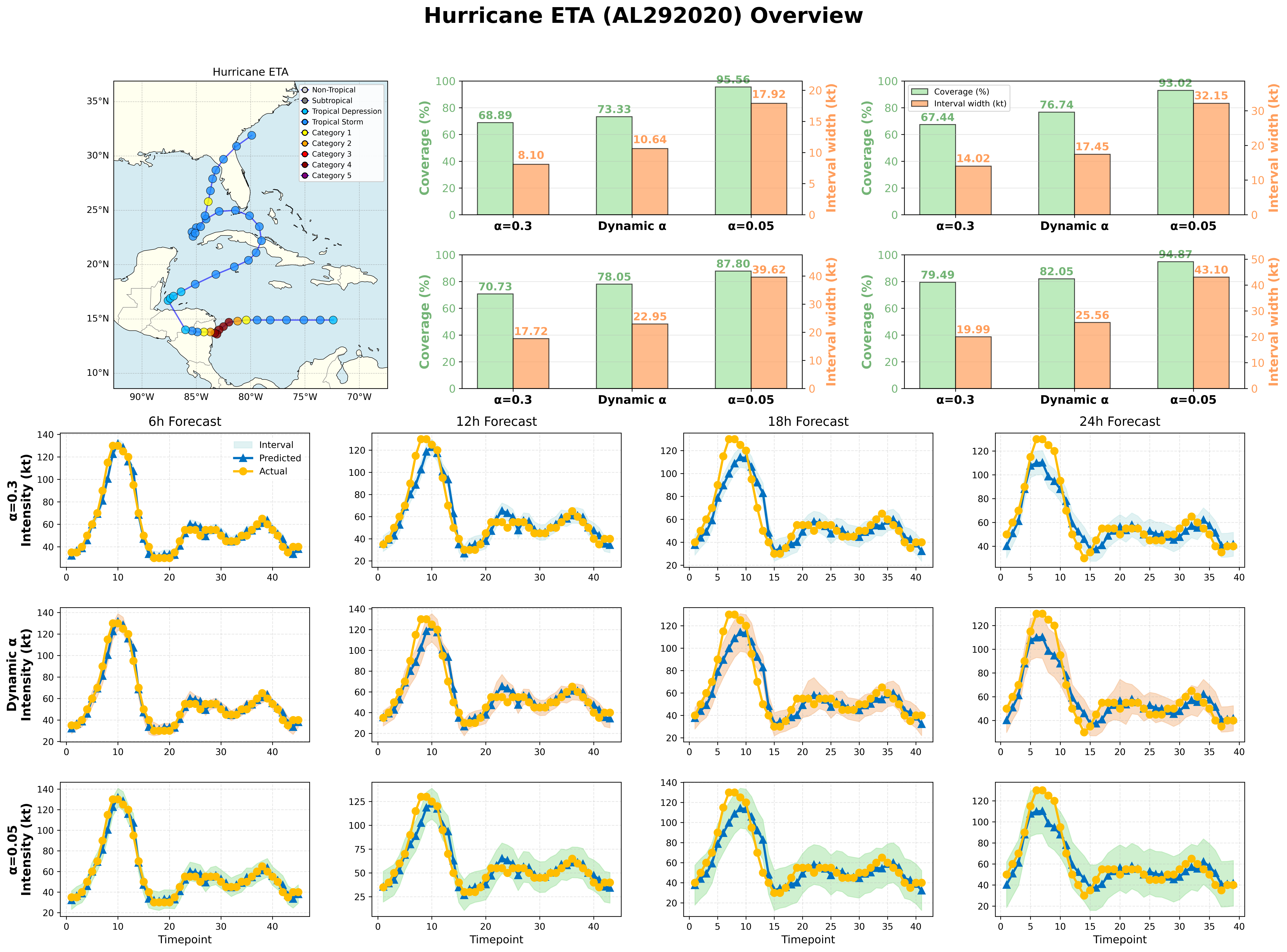}
    \caption{Prediction results of Hurricane ETA, including the observed track and comparisons of prediction interval coverage and width at 6-hour to 24-hour forecasts under fixed $\alpha = 0.3$, dynamic $\alpha$, and fixed $\alpha = 0.05$.}
    \label{fig:4}
\end{figure}
Figure 4 shows the results of an independent test of Hurricane Eta, the most intense and influential hurricane of 2020, containing forecasts from 6-hour to 24-hour. The hurricane continued to rapidly increase in intensity from 0:00 UTC to 18:00 UTC on November 2, peaked in intensity at 0:00 UTC on November 3, then rapidly weakened, and remained largely stable thereafter. Compared to alpha = 0.3: the dynamic alpha approach improves the coverage by 8.7\%-6.6\%, especially during the most intense part of the intensification, and the dynamic alpha provides wider prediction intervals, which reduces the coverage by about 10\%-15\% but the interval widths by almost 50\% compared to alpha = 0.05. During the RI phase, the model's point predictions underestimate the actual intensity, and dynamic alpha effectively compensates for this bias by widening the upper boundary, especially in the 12h and 18h predictions, where this correction is most effective.

In addition, two specific cases—Hurricane Larry and Hurricane Genevieve—were shown in the SI. Our method demonstrated consistently strong performance. In particular, for Hurricane Genevieve, the 24-hour forecast with dynamic alpha achieved 94.1\% coverage, much higher than 58.8\% with alpha = 0.3, and a narrower interval width (28.5 vs. 43.1 kt) compared to alpha = 0.05, demonstrating a good balance between accuracy and uncertainty. This demonstrates that our method effectively balances coverage and prediction interval width.Notably, both Genevieve and Larry underwent RI, and our method was able to promptly detect RI and adaptively widen the prediction intervals, resulting in improved coverage under such challenging conditions. 
 
\section{Conclusion}
In summary, our study introduces a novel and interpretable uncertainty quantification framework for tropical cyclone intensity forecasting, which, for the first time, incorporates TC classification criteria and developmental stage information into the CP process.Experiments demonstrate we move beyond static confidence bands and offer real-time adaptive forecasting tailored to the cyclone's development phase. This significantly advances the capability to quantify uncertainty in extreme weather events, marking a step toward more reliable and operationally relevant early warning systems. 

Although this study improves uncertainty quantification in TC intensity prediction, the current framework uses intensity-based stage identification, the incorporation of dynamic and thermodynamic environmental factors—such as vertical wind shear, ocean heat content, and mid-level humidity—remains a promising direction for further enhancing physical adaptability. Moreover, the Transformer-based predictive model lacks explicit physical constraints, operating largely as a “black box”; integrating physics-informed layers may enhance the physical consistency of future predictions.

\appendix
%\section{Here is a sample appendix}

%%%%%%%%%%%%%%%%%%%%%%%%%%%%%%%%%%%%%%%%%%%%%%%
% Optional Glossary, Notation or Acronym section goes here:
%
% Glossary is only allowed in Reviews of Geophysics
%  \begin{glossary}
%  \term{Term}
%   Term Definition here
%  \term{Term}
%   Term Definition here
%  \term{Term}
%   Term Definition here
%  \end{glossary}

%%%%%%%%%%%%%%%%%%%%%%%%%%%%%%%%%%%%%%%%%%%%%%%
% Acronyms
%% NOTE that acronyms in the final published version will be spelled out when used in figure captions.
%   \begin{acronyms}
%   \acro{Acronym}
%   Definition here
%   \acro{EMOS}
%   Ensemble model output statistics
%   \acro{ECMWF}
%   Centre for Medium-Range Weather Forecasts
%   \end{acronyms}

%%%%%%%%%%%%%%%%%%%%%%%%%%%%%%%%%%%%%%%%%%%%%%%
% Notation
%   \begin{notation}
%   \notation{$a+b$} Notation Definition here
%   \notation{$e=mc^2$}
%   Equation in German-born physicist Albert Einstein's theory of special
%  relativity that showed that the increased relativistic mass ($m$) of a
%  body comes from the energy of motion of the body—that is, its kinetic
%  energy ($E$)—divided by the speed of light squared ($c^2$).
%   \end{notation}

%%%%%%%%%%%%%%%%%%%%%%%%%%%%%%%%%%%%%%%%%%%%%%%
%
% DATA SECTION and ACKNOWLEDGMENTS
%
%%%%%%%%%%%%%%%%%%%%%%%%%%%%%%%%%%%%%%%%%%%%%%%
\section*{Open Research Section}
The data supporting the conclusions of this study are available from the SHIPS database, hosted by the Colorado State University, and can be accessed at \url{https://rammb2.cira.colostate.edu/research/tropical-cyclones/ships/}. The code and data supporting the results of this study will be made publicly available at \url{https://github.com/stevewinwin/Adaptive-CP-TC-Intensity} upon acceptance for publication.

\acknowledgments
The authors gratefully acknowledge the supports by National Natural Science Foundation, Basic Science Center Project, 42088101, Climate System Prediction Research Center and the Natural Science Foundation of Jiangsu Province(Grants No. BK20240700).

%%%%%%%%%%%%%%%%%%%%%%%%%%%%%%%%%%%%%%%%%%%%%%%
% REFERENCES and BIBLIOGRAPHY
%
 %\bibliography{<name of your .bib file>} don't specify the file extension
% don't specify bibliographystyle
%
%%%%%%%%%%%%%%%%%%%%%%%%%%%%%%%%%%%%%%%%%%%%%%%
\bibliography{aref}

%Reference citation instructions and examples:
%
% Please use ONLY \cite and \citeA for reference citations.
% \cite for parenthetical references
% ...as shown in recent studies (Simpson et al., 2019)
% \citeA for in-text citations
% ...Simpson et al. (2019) have shown...
%
%
%...as shown by \citeA{jskilby}.
%...as shown by \citeA{lewin76}, \citeA{carson86}, \citeA{bartoldy02}, and \citeA{rinaldi03}.
%...has been shown \cite{jskilbye}.
%...has been shown \cite{lewin76,carson86,bartoldy02,rinaldi03}.
%... \cite <i.e.>[]{lewin76,carson86,bartoldy02,rinaldi03}.
%...has been shown by \cite <e.g.,>[and others]{lewin76}.
%
% apacite uses < > for prenotes and [ ] for postnotes
% DO NOT use other cite commands (e.g., \citet, \citep, \citeyear, \nocite, \citealp, etc.).
%

\end{document}